\documentclass[amsmath,aps,12pt]{revtex4}%,twocolumn
\usepackage{amsthm,amsfonts,graphicx,verbatim}
\newcommand{\be}{\begin{equation}}
\newcommand{\ee}{\end{equation}}
\newcommand{\bea}{\begin{eqnarray}}
\newcommand{\eea}{\end{eqnarray}}

\newcommand{\la}{\langle}
\newcommand{\ra}{\rangle}

\renewcommand{\epsilon}{\varepsilon}

\begin{document}
\title{On the stability of topological phases on a lattice}
\author{Israel Klich }
\affiliation{ Department of Physics,
University of Virginia, Charlottesville, VA
22904}

\begin{abstract}
We study the stability of anyonic models on lattices to perturbations. We establish a cluster expansion for the energy of the perturbed models and use it to study the stability of the models to local perturbations. We show that the spectral gap is stable when the model is defined on a sphere, so that there is no ground state degeneracy. We then consider the toric code Hamiltonian on a torus with a class of abelian perturbations and show that it is stable when the torus directions are taken to infinity simultaneously, and is unstable when the thin torus limit is taken.
\end{abstract}
\maketitle

The presence of a gap is of crucial importance when considering the
utility of topological phases for fault tolerant quantum computation \cite{kitaev2003fault}. A gap ensures exponential decay of correlations, the stability of operations
to noise by localized perturbations and the possibility of adiabatically "dragging" anyons in order to perform gates. For a review of the ideas of topological quantum computing see e. g. \cite{nayak2008non}.   
Also, as shown by Hastings and Wen \cite{hastings2005quasiadiabatic} the ground state degeneracy and structure of topological states (i.e. the exponentially small splitting of ground state degeneracy) is retained if a gap stays open along a path in the space of Hamiltonians connecting the perturbed and non perturbed Hamiltonian. Thus, it is important to verify the existence of a gap when searching for models with non abelian excitations which can be described by simple spin systems, and are rich enough to be useful for topological quantum computing. A recent example is Fendley's model \cite{fendley2008topological}, which is presumed to exhibit non-abelian quantum excitations, however, the existence of a gap is still an open issue.
Understanding the presence and magnitude of gaps in generic Hamiltonians is also of critical importance for notion of adiabatic quantum computing  \cite{farhi2000quantum}.

In addition, temperature may play a competing role in destroying
topological stability, and also affects the scaling of entanglement entropy as shown for the toric code and related Hamiltonians  \cite{castelnovo2007entanglement,castelnovo2007topological, nussinov2008autocorrelations, iblisdir2009scaling}. %However it is important to emphasize that stability to perturbations is a substantially harder problem than understanding the role of temperature, because the Toric code is exactly solvable at any temperature (ACTUALLY THEY DO MORE). 
%The effect of temperature on the Toric Code (and other stabilizer states) was considered in several works %(Bravyi,Nussinov,Alicki...) ({\bf WHAT IS THE MAIN CONCLUSION?}). 
A way of addressing the fragility of the phase to temperature was proposed in \cite{hamma2009toric} and relies on confining the strings separating anyon excitations with a string tension terms. In \cite{hamma2009toric} such terms are generated by coupling to a bosonic field.

The presence of a gap in large quantum systems lies at the heart of our understanding of correlations and phases. %It is of fundamental interest, as indicating different
%quantum phases of materials. 
A common working assumption involved in treating many such problems is that the existence of a
spectral gap ensures the stability of a phase to small perturbations. This is indeed the case whenever considering perturbations which are smaller in operator norm than the gap, by the convergence of perturbation theory. Thus a small localized perturbation cannot do real damage to the phase. 
However, such an argument is not useful when determining the stability of a given phase. Indeed, in any physical 
realization of a Hamiltonian describing a given phase, one may expect some degree of deviation from an idealized model, which is
spread throughout the sample, and as such, may have a large norm, scaling as the volume of the system when a thermodynamic limit is taken. 

While showing that a system is gapless may sometimes be done by variational means as is, for example, done in the proof of Lieb-Schultz-Mattis theorem for spin $S=1/2$ Heisenberg model \cite{lieb1961two} (and its generalization to higher dimensions \cite{hastings2004lieb}), establishing the presence of a gap may be substantially more involved, and cannot be addressed in this way satisfactory, as is seen from the closely related but much harder problem of establishing that a gap is present in the spin $S=1$ Heisenberg chain (known as Haldane's conjecture).
%Indeed, recently a million dollar
%prize has been proposed by the Clay institute for a proof that Yang-Mills theory is gapped \cite{MillenniumYangMills}.

In view of the importance of a gap for the topological quantum computing idea, this paper is concerned with establishing a stability theory for such systems to quantum perturbations. 

\begin{figure}
\includegraphics[scale=0.2]{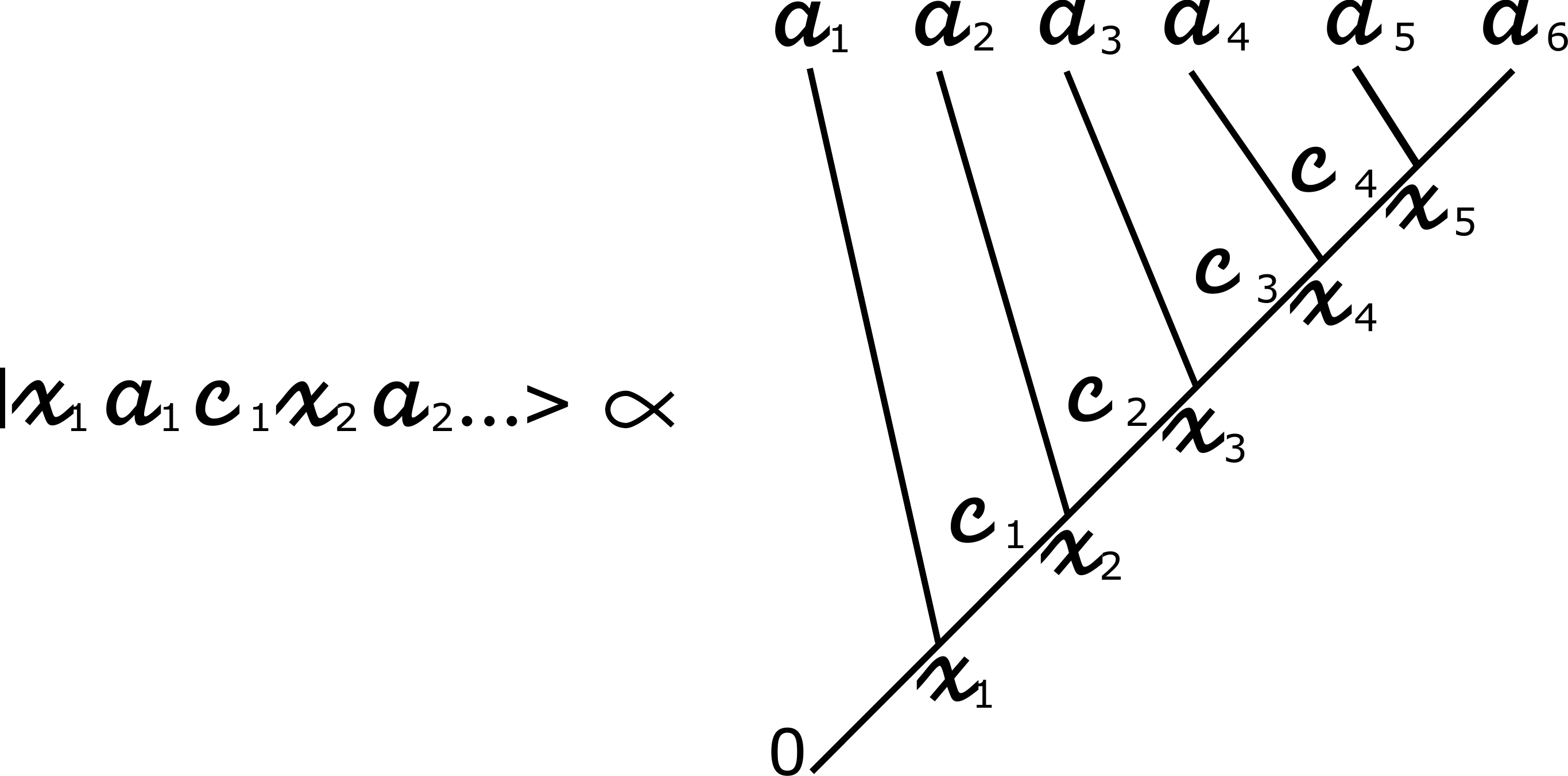}\caption{A basis element of the anyon lattice.} \label{anyonbasis}
\end{figure}

Here we consider a class of Hamiltonians describing topological anyonic models on a lattice.
An anyonic lattice model is a lattice $\Lambda$ together with a finite set of anyonic charges ${\cal C}$ and fusion and braiding rules (which constitute a "unitary braided tensor category"). The Hilbert space ${\cal H}(\Lambda)$ of the theory is spanned by fusion diagrams. These describe the charges $a_i\in{\cal C}$ occupying sites $i\in\Lambda$ and the splitting process that was involved in creating them.  
We can write a state of the system in the form:
\begin{eqnarray}\label{basis state}
|\chi_1,a_1,c_1,\chi_2,a_2,c_2,...\chi_{N-1},a_{N-1},a_N\ra
\end{eqnarray}
Here one chooses some ordering of the lattice points, so that $a_1$ is the charge at site $1$, $a_2$ the charge at site $2$ and so on. These states may be conveniently represented by splitting diagrams as in Fig. \ref{anyonbasis}.  The proportionality sign indicates that there is scalar factor relating the diagram and the normalized state. This factor is usually expressed in terms of the quantum dimension of the appearing charges, and is used to ensure the isotopy invariance of the various braiding and fusion operations that are carried out when expressed through these diagrams.
The diagrams describe the generation of states from the vacuum. First, creation of a pair $a_1,c_1=\bar{a_1}$ in a channel $\chi_1$, then splitting $ {c_1}\rightarrow a_2\chi_2$ through an allowed fusion channel with $\chi_2$ indexing the degeneracy. We continue this way until the charges at all points on the lattice are specified  (note that some of the $a_i$s may be vacuum, denoted $a_i={\bf 1}$, corresponding to no charge at location $i$). We also assume that only a finite number of different charges are allowed to sit at a given site, and that these states are normalized. 

The fusion diagrams may be thought of as word-lines of particles which are created from the vacuum state. 
Closed diagrams correspond to processes which start in the vacuum state and return to it, and so correspond to a vacuum to vacuum amplitude, or a vacuum expectation value.
The detailed algebraic structure of these theories was discovered and studied in \cite{moore1989classical, witten1989quantum,fredenhagen1989superselection,frohlich1990braid} (see also \cite{turaev1994quantum,bakalov2001lectures} for a purely mathematical treatment). A useful summary of the essentials can be found in \cite{kitaev2006anyons,BondersonThesis}, but a detailed treatment is not essential for the present work. Here the only important feature we use is that the vacuum expectation value of two braids which are unlinked is assumed to be factorizable: i.e. if the world lines consists of two closed braids $c_1$ and $c_2$ which are unlinked (and for systems on torus, not completing a full cycle of the torus) then $\la c_1\bigcup c_2\ra=\la c_1\ra\la c_2\ra$, where the expectation values are taken in a ground state of the system, i.e. a state containing no quasi particles. Thus the results of the present work may be extended to non-anyonic models once such a property can be established.  

Prime examples for anyonic lattices rise in the description of certain lattice spin Hamiltonians such as Kitaev's toric code \cite{kitaev2003fault} and the Levin-Wen models \cite{levin2005string}. The latter supplies an explicit prescription for generating all discrete gauge theories and all doubled Chern-Simons theories in (2+1) dimensions. Another way of getting anyonic lattices is to consider at the outset an array of quasi particles with anyonic properties (such as the ones arising in fractional quantum hall effect), and assuming that their interaction can be completely specified by braiding and fusing operations. An explicit example for this is the array of Fibonacci anyons known as the "golden chain" and its generalizations  \cite{feiguin2007interacting,gils2009topology}. 

% mention that the particles might exist, notably in FQHE, cite parsa \cite{bonderson2006detecting,stern2006proposed}? and wiley?
It is worth pointing out that even if the model is actually an effective description of a spin model, the particular lattice over which the corresponding anyonic lattice model is defined is not necessarily equivalent to the original lattice of spins, since particles may be associated with plaquettes of the original spin lattice and not with individual spins.

As our unperturbed Hamiltonian, we take:
\begin{eqnarray}
H_0=\sum_{i\in\Lambda} h_i
\end{eqnarray}
such that:
\begin{eqnarray}
h_i|\chi_1,a_1,c_1,\chi_2,a_2,c_2,...\chi_{N-1},a_{N-1},a_N\ra=h(a_i)|\chi_1,a_1,c_1,\chi_2,a_2,c_2,...\chi_{N-1},a_{N-1},a_N\ra
\end{eqnarray}
where $h$ is a function $h:{\cal C}\rightarrow \mathbb{R}$ such that $h({\bf 1})=0$ and $h(a_i)\geq h_0\geq 1$ for any $a_i\neq {\bf 1}$. It is clear from the definition that $[h_i,h_j]=0$. We denote $h_m=\max_{\cal C} h(a)$.

\begin{figure}
\includegraphics[scale=0.2]{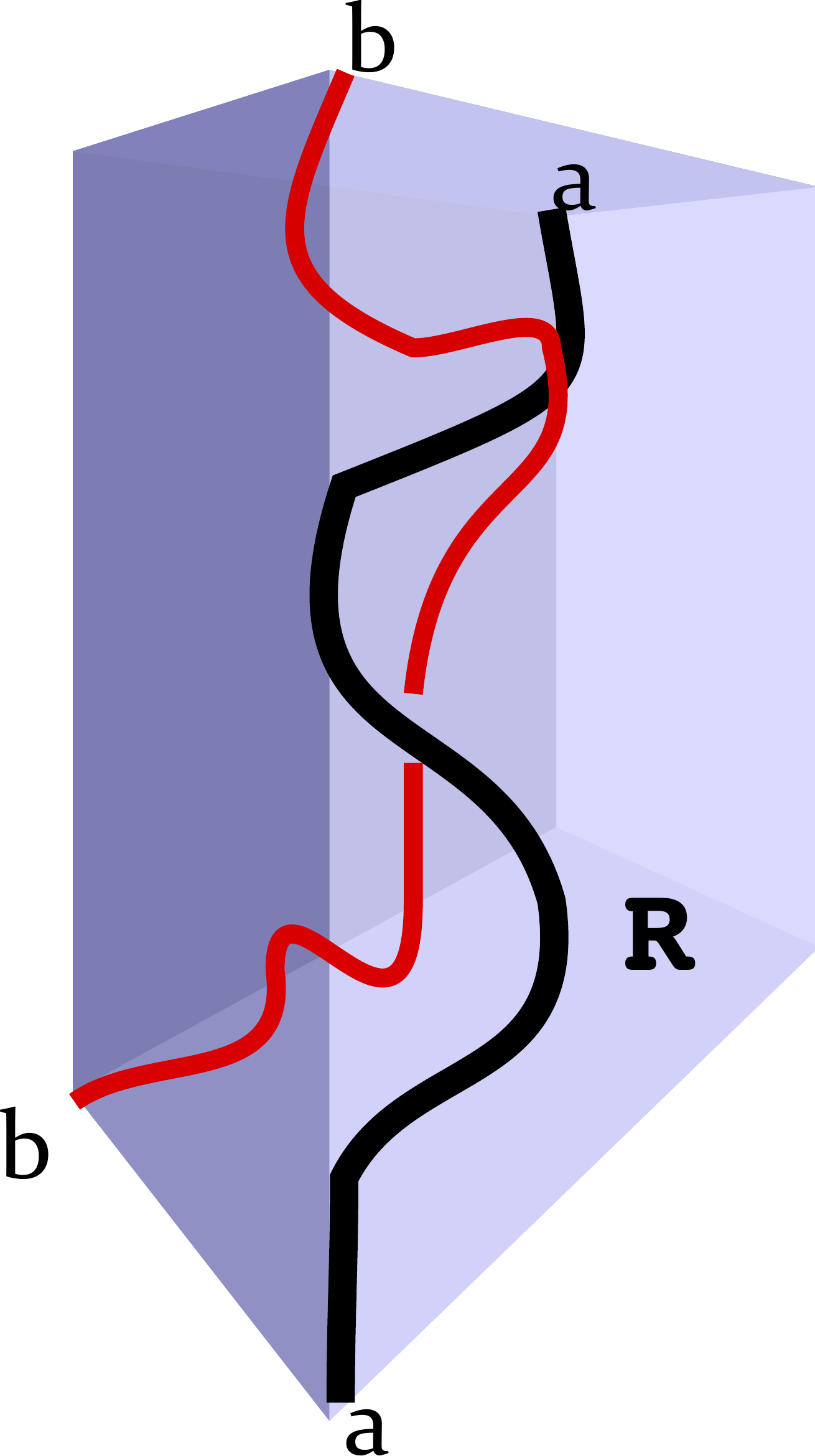}\caption{A local perturbation $V_i$ only allows braids and fusion in $R_i\times [0,1]$, so that the end points sit on lattice cites.} \label{hull of R}
\end{figure}

We now introduce the perturbations $V_i$ to the system. Admissible $V_i$s have a finite range and act in a compact simply connected region $R_i\subset {\mathbb{R}}^2$, so that a $V_i$ represents an action on a given state by combinations of braiding and fusing only charges $a_j$ on sites $j\in R_i$, and such that the braid/fusion actions occur entirely within  $R_i\times [0,1]$ and the initial and final points of the picture lie on lattice points belonging to $R_i$ (Fig. \ref{hull of R}). 

\begin{figure}
\includegraphics[scale=0.2]{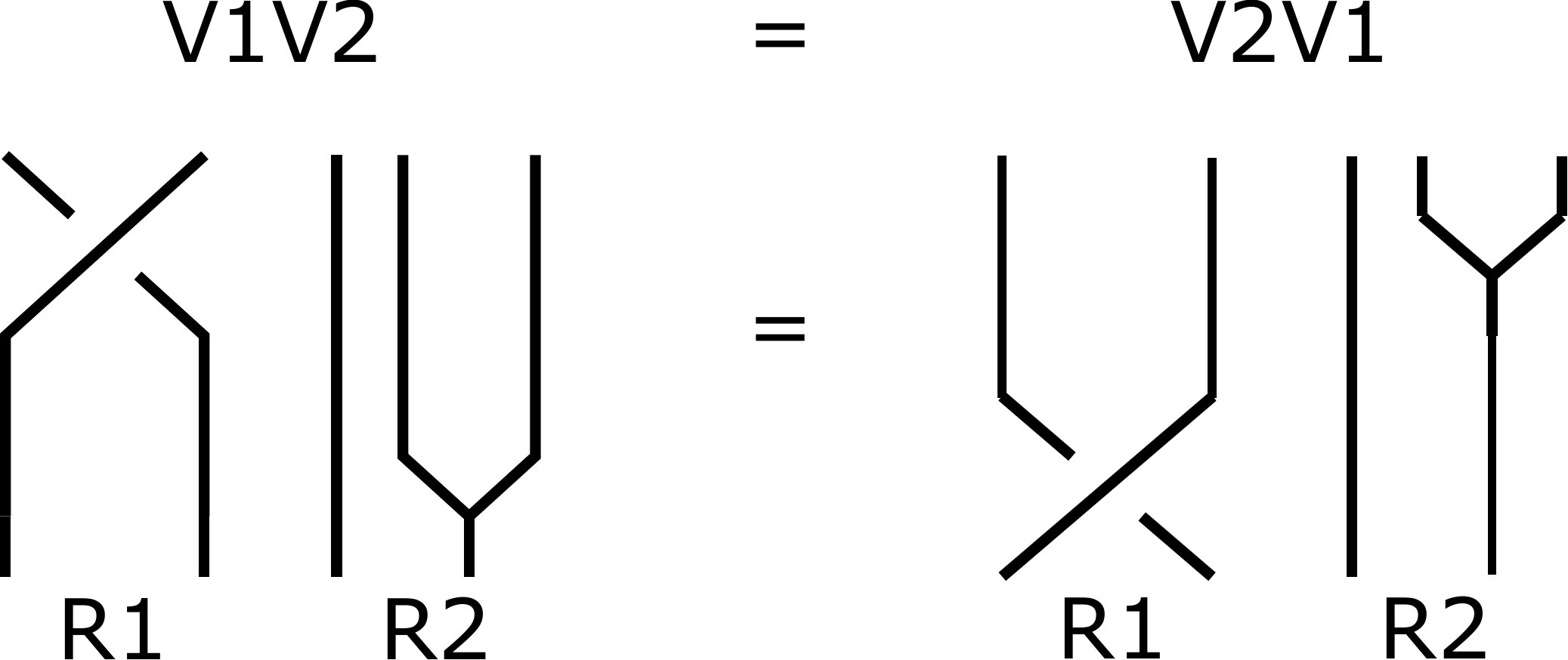}\caption{$V_i$s associated with disjoint $R_i$'s are commuting.} \label{commuting}
\end{figure}

In general, even for neighboring sites $i,j$, $R_i\cap R_j\neq\emptyset$, and the $V_i$'s do not commute. However if $R_i\cap R_j=\emptyset$ then $[V_i,V_j]=0$, this can be verified directly from action on basis states (or for models arising from a spin lattice, by recalling that all local operators can be written as local products of spin operators), and illustrated in Fig. \ref{commuting}. 

Our first result is the following:\\

%\begin{thm}
\underline{{\it Stability to quantum perturbations:}}\\
Assume that the $V_i$s are bounded:
\begin{eqnarray}
\la\psi| V_i|\psi\ra\leq 1  \,\,\,\,\, \forall \psi \in {\cal H}(\Lambda),
\end{eqnarray}
 the ground state of $H_0$ is unique and the system is defined on a sphere (i.e. there are no non-trivial cycles). Then there exists a $\beta_0>0$ which is independent of the size of the lattice (but depends on the dimension and coordination number) such that for any $\beta<\beta_0$, $H_0+\beta \sum_{i\in\Lambda}V$ has a unique ground state and spectral gap $\gamma>c(\beta)$, where $c(\beta)$  is independent of lattice size.
% \end{thm}

This part of the paper establishes the stability of the quantum phase described by $H_0$ to quantum perturbations, when $H_0$ has a unique ground state and may be viewed as a special case of \cite{datta1996low}, where stability of classical models was studied using cluster expansions and the possible applicability to anyonic systems was taken into account, albeit without a concrete framework, since the importance of anyonic systems on lattices models and examples such as Kitaev's model and Levin-Wen models where absent. 

Next, we consider a situation of a system having a degenerate ground state, which is encountered in anyonic models, when placing the lattice on a manifold with a non-trivial genus. 
 One of the simplest, and best known examples for a spin lattice Hamiltonian which yields an anyonic model is Kitaev's toric code \cite{kitaev2003fault}.  This model was first proposed to be used as a quantum memory, by storing the quantum information in the subspace spanned by the ground states (the degeneracy of the ground states is $4^g$ for a compact orientable surface of genus $g$). Here our motivation is to prove explicitly that Kitaev's argument for the stability of this memory does work. The argument is very simple: mixing between the ground states has to involve processes whereby a particle-antiparticle is created one of them tunnels around the torus and then they are annihilated, this cannot happen below $L$th order perturbation theory (where $L$ is the length of shortest of the torus cycles), and so one expects an exponentially small splitting of the ground states $o(e^{-a L})$ for some $a$. Much of what follows, is a way to establish this argument in a more explicit (and rigorous) way.
 
 Let us briefly review the toric code model. The system is
composed of spins on a the edges of a square lattice. The Hamiltonian is given by :
\begin{eqnarray}\label{toric code hamiltonian}
H_{TC}=\sum_{s} P_{+,s}+\sum_{p}P_{\square,p}
\end{eqnarray}
where $P_{+,s}={1\over 2}(1-\prod_{j\in{\rm star}(s)}\sigma_j^x)$. Here ${\rm star}{(s)}$ is the set of four edges connected to a given vertex $s$ on the lattice.
$P_{+,s}$ projects on states where odd number of spins in a star$(s)$ point in
the $x$ direction. Similarly $P_{\square,p}={1\over 2}(1-\prod_{j\in{\rm
boundary}(p)}\sigma_j^z)$ projects on states where an odd number of spins on the edges of a square 
plaquette $p$ point in the $z$ direction.
One can easily verify that all terms in $H_{TC}$ mutually commute, so it can be diagonalized simultaneously together with all the projectors $P_{+,s}$ and $P_{\square,p}$. The ground states can be thought of as states simultaneously satisfying the constraints $P_{+,s}=0$ and $P_{\square,p}=0$ for all stars and plaquettes.

In this model there are two kinds of particles, denoted $x,z$. The $x$ particles (termed "magnetic" in \cite{kitaev2003fault}) are introduced by applying 
\begin{eqnarray}
X_C=\prod_{\alpha\in C}\sigma^x_{\alpha}
\end{eqnarray}
where $C$ is some set of edges of the square lattice to a ground state. Similarly $z$ particles ("electric") are introduced by applying 
\begin{eqnarray}
Z_C=\prod_{\alpha\in C}\sigma^z_{\alpha}
\end{eqnarray}
to a ground state. In this system the $z$ particles are associated with vertices touched by odd number of edges in $C$, while the $x$ particles are associated with plaquettes surrounded by odd number of edges in $C$.
Thus, for example given an edge $\alpha=(i,j)$ an application of  $\sigma^{z}_{\alpha}$ to the ground state creates a pair of violations of the constraints $P_{+,i}=0$ and $P_{+,j}=0$ and costs energy penalty 2. The set of charges contains now $1,x,z$, and any state in the system can be generated by successive generation of particles from the ground state, as in \eqref{basis state}. Since the model is abelian, there is no degeneracy of the fusion channels and so the $\chi$ labels are all trivial. 

When the model is placed on a sphere, the ground state is non-degenerate and so the stability theorem described above applies. We conclude that the toric code on the sphere is stable to perturbations.

Next, we consider the toric code on a torus, that is, our lattice of spins is now assumed to have periodic boundary condition, and we identify the sites $(i,j)=(i+L_1,j)$ and $(i,j)=(i,j+L_2)$. 
The toric code has now four ground states, which may be characterized by applying loop operators to torus cycles.

To understand what may happen under this degeneracy we proceed to study a particular kind of perturbations. The perturbations we consider are of the form:
\begin{eqnarray}
V=\sum V_i
\end{eqnarray}
each $V_i$ is composed solely out of sums and products of $\sigma_z$ operators, and so can only create $z$-charges (electric).
(Alternatively one may choose $V_i$ to be composed of $\sigma_x$
operators and create "magnetic" charges).  In 
\cite{trebst2007breakdown} a particular case of such a perturbation was shown numerically to preserve the topological phase. 

Our main results for this model are:

1) One can completely destroy the gap if $L_2$ is kept fixed while $L_1\rightarrow\infty$.

2) If $\log(L_1)/L_2$ and $\log(L_2)/L_1$ are kept finite when the limit $L_2,L_2\rightarrow\infty$ then $H_{TC}$ is stable to the $V_i$ perturbations.
\\
This is a preliminary result for a more general treatment of general anyonic systems on a torus (without the restriction on the type of perturbation).

\section{Proof of stability}
In this section we use techniques developed for the study of classical models and their perturbations and adapt them to the present setting. A {\it classical model} is a local Hamiltonian model defined on a tensor product Hilbert space (e.g. a spin lattice) in which the local terms are diagonalized simultaneously by a basis of product states. Indeed, while nontrivial topological Hamiltonians are never classical models, the particular form of the unperturbed Hamiltonians which we use, i.e. a commuting sum of local constraints, has much of the structure of a classical Hamiltonian.
The stability of classical phases to temperature and quantum perturbations has been a rich field of study with well established results in the frame work of Pirogov-Sinai theory and it's subsequent refinements see e.g.\cite{pirogov1976phase,zahradnik1984alternate,borgs1989unified,borgs1996low,datta1996low}. For classical Hamiltonians, stability of phases in the case of degenerate ground states usually depends on the Peierls condition.

In the present paper we consider the correction to ground state energy and gap of topological  lattice Hamiltonians  by studying 
\begin{eqnarray}\label{ZN}
Z(N)=\la \alpha|e^{-N (H_0+V)}|\alpha \ra
\end{eqnarray}
where $|\alpha \ra$ is a ground state of the system (possibly one of a set of ground states). If there is a gap, it follows that as $N\rightarrow\infty$ 
\begin{eqnarray}\label{asymptotic Z}
  \log(Z)=c_{\alpha}-E_{\alpha} N+o(e^{-\gamma_{\alpha} N})
\end{eqnarray}
where $E_{\alpha}$ is the perturbed ground state energy in the cyclic subspace generated by applications of $H$ on $|0 \ra$, and $\gamma_{\alpha}$ is the gap above this ground state, in this cyclic space.

%The $z$ particles
%commute, rendering the cyclic subspace generated by applying
%$H_0+V$ to the ground state of the $H_0$ be described by localized
%bosons. 

It is convenient to compute $\log(Z_N)$ using cluster expansions. These have been extremely useful for the task of studying stability of classical models. However, the main difficulty in applying these considerations here, lies in the non-local nature of anyonic systems. This arises in two ways: 1) large clusters may be correlated with arbitrarily far other large clusters if they span non trivial loops around the torus, and as such cannot satisfy the usual factorizability properties needed to establish the convergence of the cluster expansion, and 2) clusters which are disjoint but {\it linked} are not factorizable. In this section we consider systems defined on a sphere, so that correlation between arbitrarily far large clusters is not an issue and ground state degeneracy is not an issue. A system on a torus exhibiting a degenerate ground state and correlations between large clusters will be studied in the next section.

There are a variety of ways of getting cluster expansions. Perhaps, the most straightforward one is to write a Duhamel expansion of the partition function as is done in many works. 
However, here we chose to follow closely an elegant and more compact way by Yarotsky \cite{yarotsky2006ground}, in order to get such an expansion, with the desired convergence properties. 

Given a set of vertices $I \subset \Lambda$, we associate to it a projection operator $P_I=\prod_{s\in I}P_{s}$, which projects on the states in which no charge other then the vacuum ${\bf 1}$ sits in $I$, and the operator $Q=\prod_{s\in I}P^{\perp}_{s}$, which project on a state where all vertices in $I$ carry a non-trivial charge. Using the inclusion-exclusion principle $Z$ may be compactly written as:
\begin{eqnarray}\label{expansion}
  Z=\sum_{c}\omega(c).
\end{eqnarray}
Here $c=\cup_{k=1}^N\{I_k,J_k\}$, $I\subset\Lambda$, $J\subset\Lambda\backslash\bar{I}$, is a configuration on $N\times \Omega$,

\begin{eqnarray}\label{omega_c}
  \omega(c)=\la 0| \prod_{k=1}^N T_{I_k,J_k}| 0\ra
\end{eqnarray}
and 
\begin{eqnarray}
T_{I,J}=\sum_{L\subset I}{-1}^{|I|-|L|}e^{-\sum_{i\notin \bar{I}}h_i}e^{-H_L}Q_{J}P_{\Lambda \backslash\bar{I} \backslash J}. 
\end{eqnarray}
 Here 
$\bar{I}=\bigcup_{i\in I}\bigcup_{j\,s.t.\,i\in R_j}R_j$, i.e. the set of points $i$'s in $\Lambda$ s.t. the site is affected by a perturbation in $I$ and $H_L=\sum_{i\in \bar{I}}h_i+\sum_{i\in L}V_i$. 

To obtain the expansion \eqref{expansion} we have used that for any $I\subset\Lambda$
\begin{eqnarray}
\sum_{J\subset\Lambda\backslash \bar{I}}Q_JP_{\Lambda \backslash\bar{I} \backslash J}={\rm Id}
\end{eqnarray}
where Id is the identity operator, and the inclusion exclusion principle through:
\begin{eqnarray}
e^{-H}=\sum_{I\subset\Lambda}\sum_{L\subset I}{-1}^{|I|-|L|}e^{-\sum_{i\notin \bar{I}}h_i}e^{-H_L}
\end{eqnarray}
Therefore:
\begin{eqnarray}
e^{-NH}=\sum_{c}\prod_{k=1}^NT_{I_k,J_k}
\end{eqnarray}
The expectation value in $\omega(c)$ is taken in the vacuum state. 

A configuration $c$ can be identified with a set in $\Lambda\times(1,..N)$. We define the support of a configuration by thickening the configuration in the time direction: define $\tau{(i,k)}=(i,k+1)$, then $supp(c)=\bar{c}\bigcup( \tau \bar{c})\bigcup (\tau^{-1}\bar{ c})$ in addition, we add a line for any two points in $c$ which are next nearest neighbors on the lattice at a given "time slice",  or are nearest neighbors in the "time direction" (Note that two different configurations can have the same support, depending on the  interaction, for example if configuration (1) is the entire $\Lambda\times(1,..N)$ and configuration (2) is $\Lambda\times(1,..N)$ without one site they will have the same support the entire $\Lambda\times(1,..N)$.). We denote the volume of a configuration $|c|=\{\#$ of points in $c\}$.

The key property to evaluating $Z_N$ is factorizability. For classical models, one usually shows that if $c=c_1\bigcup c_2$, and $supp(c_1)\cap supp(c_2)=\emptyset$ then: $\omega(c)=\omega(c_1)\omega(c_2)$.

For anyonic models this is false. However we have a weaker kind of factorization property: If $c=c_1\bigcup c_2$, and $supp(c_1)$ and $supp(c_2)$ are non-linked, then: $\omega(c)=\omega(c_1)\omega(c_2)$.
To establish this property, we can take two configurations $c_1$ and $c_2$, and expand the $T$ operators involved in \eqref{omega_c} in terms of the perturbations $V_I$s. The action of those on the ground state, in turn, can be expressed as a superposition of braid/fusion diagrams. It is clear that each term in such an expansion of $c_1$ and $c_2$ consists of a superposition braids, such that any pair of braids, one belonging to $c_1$ and one to $c_2$ are unlinked. By the axioms of the anyonic models we have that the vacuum expectation values of any such a pair factors. Thus we say that $c_1$ and $c_2$ are compatible if their supports are not linked.

Factorizability allows us to use the linked clusters in the following way. Write \cite{gallavotti1973some,simon1993statistical}:
\begin{eqnarray}
\log Z_N=\log\sum_{irreducible\& unlinked\,\,\,c_1,..c_n}\prod \omega(c_k)=\sum_{X} \omega(X)
\end{eqnarray}
where $X$ is a collection of polymers $c_k$ with multiplicities $n_k$, and,
\begin{eqnarray}\label{omegaX}
\omega(X)=\prod{\la c_k \ra^{n_k}\over n_k!}\sum_{G_1\lhd G(X)}(-1)^{l(G_1)}.
\end{eqnarray}
Here $G(X)$ is a graph obtained from $X$, which consists of $\sum n_k$ vertices, where there are $n_k$ vertices which are associated to each $c_k$ in $X$ . For two vertices associated with $k_1$ and $k_2$, say, there is an edge if and only if $c_{k1}$ and $c_{k2}$ are not compatible (i.e. their support intersects or links). The sum on the right is over complete connected subgraphs: $G_1\lhd G(X)$ i.e. $G_1$ is a subgraph of $G(X)$, such that $G_1$ is a connected and contains all vertices of $G(X)$. In particular, if $G(X)$ may be separated into two disconnected component, then there are no such $G_1$, and so $\omega(X)=0$, which justifies the name "linked cluster expansion".

To make this expansion useful we now need to establish two properties: \\
1) Exponential bound on the number of clusters of given size and \\ 2) Exponential decay of $\omega(c)$ with the volume of $c$.

\subsubsection{Exponential bound on the number of linked objects} 
\begin{figure}
\includegraphics[scale=0.2]
{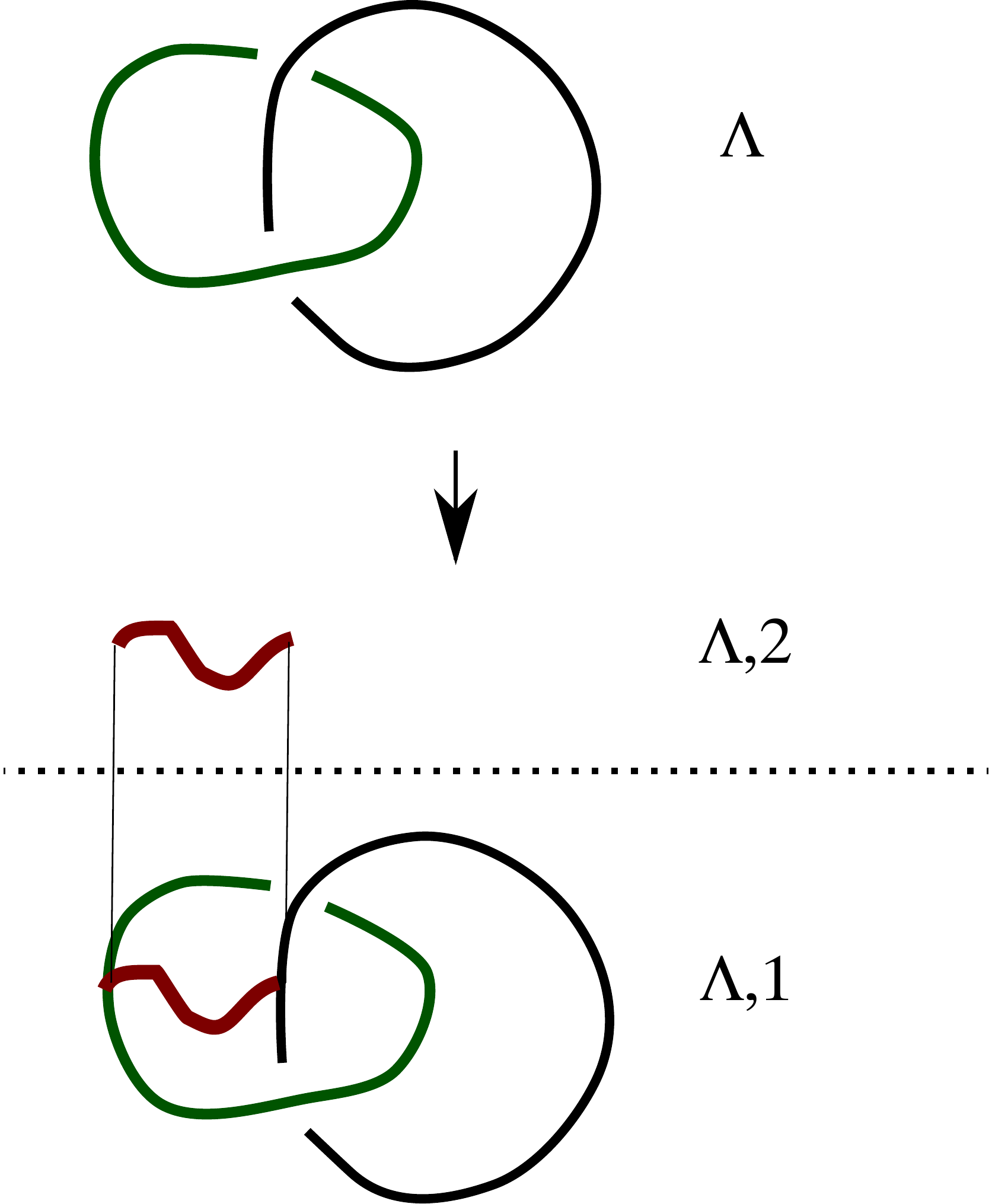}\caption{For each {\it linked} configuration in the lattice $\Lambda$ we can choose a corresponding {\it connected} configuration of larger size in $\Lambda_1\bigcup \Lambda_2$. No two configurations in $\Lambda$ of length $L$ correspond to the same configuration in $\Lambda_1\bigcup \Lambda_2$} \label{linkedcounting}
\end{figure}

First, we prove that the number linked clusters of given volume $L$ is bounded by $\nu^{L}$ for some $\nu$.  This can be established in several ways but we use the following simple argument: Consider the doubled lattice $(\Lambda\times N) \times \{1,2\}$. We can turn any linked configuration in $((\Lambda\times N),1)$ into a unique connected configuration by adding segments of total length smaller then the linear size of the configuration (Fig. \ref{linkedcounting}) (since the distance between any two components which are linked to each other is bounded by their size). We then add this segments in the $((\Lambda\times N),2)$ (one can think of adding it in "parallel") and connect them at some point to the original added segment as illustrated in Fig. \ref{linkedcounting}. The volume of the added segments is then at most $L+N$ where $N$ is the number of connection points, therefore the size of the connected configuration in $(\Lambda\times N)\times \mathbb{Z}_2$ is bounded by $3L$ (see Fig. \ref{linkedcounting}). Since this mapping into the set of connected clusters in the doubled lattice is injective, we conclude that the number of linked clusters of size $L$ is bounded by the number of {\it connected} clusters of size $3L$ on the doubled lattice. The latter grows like ${\nu}^L$ for some constant $\nu$ which depends on the dimension and connectivity of $\Lambda$ \cite{simon1993statistical}.

\subsubsection{Exponential bound on value of clusters}
Here we repeat the argument in \cite{yarotsky2006ground}: Consider
\begin{eqnarray}
T_{I,J}=\sum_{L \subset I}{-1}^{|I|-|L|}e^{-\sum_{i\notin \bar{I}}h_i}e^{-H_L(z_1,..z_L)}Q_{J}P_{\Lambda\backslash\bar{I} \backslash J}
\end{eqnarray}
where 
$H_L(z_1,..z_L)=\sum_{i\in\bar{I}}h_i+\sum_{i\in L} z_iV_i$. Considered as a function of the $z_i$'s, 
\begin{eqnarray}\label{F}
F_{\psi}(z_1,..z_n)=\la\psi|\sum_{L\subset I}{-1}^{|I|-|L|}e^{-H_L(z_1,..z_L)}|\psi\ra
\end{eqnarray}
is analytic for any $\psi$. Note that if $|z_1|,|z_2|,..\leq 1$, 
then for any $\psi$, $\la\psi|H_L(z_1,..z_L)|\psi\ra\leq h_m|\bar{I}|+\beta|I|\leq c |I|$ for some $c$ (independent of $\beta$), where $c> 1$ is determined by the range of the $V_i$s and $h_m$.
Therefore $F(z_1,..z_n)\leq 2^{|I|}e^{|{I}|c}$ and note that for any $i\in I$, $F_{\psi}(z_1,z_2,..z_{i-1},0,z_{i+1},z_n)=0$, since in this case the contribution from any set $L\subset I \backslash i$ to \eqref{F}, is exactly cancelled by the contribution to \eqref{F} from $L\bigcup i$. By the Schwartz lemma we have that $F_{\psi}(z_1,..z_n)\leq |z_1|2^{|I|}e^{|I|c}$, for $|z_1|<\beta$. It follows that $F_{\psi}(z_1,..z_n)\leq \beta 2^{|I|}e^{|I|c}$. In the same way, $F_{\psi}(z_1,0,z_3,...)=0$, therefore again by Schwartz lemma $F_{\psi}(z_1,..z_n)\leq |z_2| \beta 2^{|I|}e^{|I|c}$, taking $|z_2|<\beta$ and repeating this argument we find that:
\begin{eqnarray}
||\sum_{L\subset I}{-1}^{|I|-|L|}e^{-H_L(z_1,..z_L)} ||\leq \beta^{|I|} 2^{|I|}e^{|I|c}
\end{eqnarray}

We are left with evaluating $||e^{-\sum_{i\notin \bar{I}}h_i}Q_{J}P_{\Lambda\backslash\bar{I}}||$, but it is easy to see that each $e^{-h_i}$ acting on a state projected by $Q_J$ is bounded by $e^{-h_0}$, and so we conclude that: $||T_{I,J}||\leq  (2\beta e^c)^{|I|}e^{-|J|h_0}$. For any desired $\epsilon$ picking $\beta$ small enough and $h_0$ large enough will ensure that $|\omega(c)|\leq {\epsilon}^{|supp(c)|}$.

Having established these facts we can proceed by the usual means of cluster theory: The exponential bound on the weight of clusters together with the exponential bound on the number of clusters imply the absolute convergence of the linked cluster expansions if  $\epsilon$ is small enough. We identify the correction to the energy, and the gap by identifying the terms in the asymptotic expression \eqref{asymptotic Z}. In the following, we sum over clusters which are defined on polymers on $\Lambda\times \infty$, note that clusters of time length $l(X)>N$ cancel out when adding the terms below to get $\log Z_N$ in \eqref{asymptotic Z}.
\begin{eqnarray}
E_0=\sum_{{X{\rm connected\,clusters\,starting\,at\,t=0}}}\omega(X)
\end{eqnarray}
The $o(e^{-N\gamma})$ term is:
\begin{eqnarray}\label{gap term}
\sum_{{X{\rm connected\,clusters\,starting\,at\,t=0\,with\, l(X)>N}}}(l(X)-N)\omega(X)
\end{eqnarray}
that this term is $o(e^{-N\gamma})$ is a consequence of summing over clusters with volume larger then $N$, together with the exponential bounds on the number of clusters and their expectation values (The relation between $\gamma$ and $\epsilon,\nu$ is described in \cite{simon1993statistical}). Finally, the 
constant term is:
\begin{eqnarray}
C=-\sum_{{X{\rm connected\,clusters\,starting\,at\,t=0}}} l(X)\omega(X)
\end{eqnarray}
In general, $E_0$, $C$ depend on the lattice $\Lambda$ and its volume. However, the exponent $\gamma$ is obtained irrespective of the size of $\Lambda$, therefore, the gap also holds when taking the thermodynamic limit.

To conclude we remark that the above treatment holds only in the cyclic subspace generated by applications of $H$ to the ground state. The treatment can be extended to the entire Hilbert space, by slightly perturbing the initial state and computing $\la 0+v|e^{-NH}|0+v\ra$ and observing the changes to the cluster expansion. The resulting additional clusters are placed in the beginning or the end of $\omega(c)$ and it is clear that as such contribute as $C_1+o(e^{-\gamma N})$ to $\log Z_N$ and do not ruin the gap \cite{yarotsky2006ground}.

\section{Toric code on a torus}

We now turn to consider in more detail the toric code Hamiltonian $H_{TC}$ defined in Eq. \eqref{toric code hamiltonian}.

The four ground states of $H_{TC}$ on a torus may be characterized by applying a $z$-loop operators to the torus cycles. These correspond to creating a particle antiparticle pair and then transporting the particle around the torus and annihilating it with the antiparticle.
Let us denote by $T_1$ and $T_2$ the two loop operators (one can choose, for example $T_1=\prod_{i=1}^{L_1}\sigma^z_{i1}$ and $T_2=\prod_{i=1}^{L_2}\sigma^z_{1i}$). 

Here we deal with perturbations $V_i$ which are functions of $\sigma^z$ operators alone.
Note that $ T_1$  and $T_2$ commute with $H_{TC}$ and all of the $V_i$ operators. 
The ground states can be chosen as eigenvectors of the $T_1$ and $T_2$. Since $T_i^2=1$, their eigenvalues are $\pm 1$, we can build the projectors: ${1\over 2}(1\pm T_1)$ and ${1\over 2}(1\pm T_2)$ which project on the symmetric and anti-symmetric subspaces.

Let us denote $|\sigma \mu\ra$ where $\sigma,\mu=\pm$ the ground state of $H_{TC}$ such that $T_1|\sigma \mu\ra=\sigma|\sigma \mu\ra$ and $T_2|\sigma \mu\ra=\mu|\sigma \mu\ra$. We stress again that any application of $\sigma_z$ operators cannot mix between these states, therefore, the cyclic Hilbert space generated by applications of $\sigma_z$ to these states is of the structure $\oplus_{\sigma=\pm,\mu=\pm} {\cal H}_{\sigma \mu}$, where 
${\cal H}_{\sigma \mu}=span\{{\rm products\,\ of\,\,} V_I {\rm \,\,applied\,\, to \,\,} |\sigma,\mu\ra\}$

\subsection{Instability of thin tori}
An idealized description of Fractional Quantum Hall wave functions, which has been has been very successful is the "thin torus limit" \cite{bergholtz2005half,bergholtz2006pfaffian,seidel2006abelian,ardonne2009domain}. 
In this situation one considers one of the directions of the torus infinitely long compared to the other. This yields a simple way of understanding the fusion rules and their application for various quantum hall
states by considering domain walls in a quasi 1d object.

Here we show that while the thin-torus limit is convenient from an algebraic/topological point of view, it also supplies an example of what can go wrong from the stability point of view: infinitesimal perturbations may ruin topological order if the ratio of the torus cycles is taken to infinity too fast.

%\begin{figure}
%\includegraphics[scale=0.4]
%{ThinTorusPerturbed.pdf}\caption{The support of the function
%$f_A$} \label{supportF}
%\end{figure}

%\subsection{Very very thin}
For the sake of exposition we look at the extreme situation where the torus has a small radius in one direction, so that we may consider a loop operator around this direction as "local". For thicker tori, loops wills arise at higher order of perturbation theory and one may show that an analogous situation to the one considered here happens. Thus, consider the perturbed toric code Hamiltonian:

\begin{eqnarray}
H_{TC}+{1\over L_1}\sum l_m
\end{eqnarray}
where $l_m=\prod_{j=1}^{L_2} \sigma^z_{j,m}$.

We note that $[l_m,H_{TC}]=0$, and that in the thermodynamic limit $L_1\rightarrow \infty$ the perturbation is locally infinitesimally small.

The ground state of this system is easily computed. Note that $l_m|\sigma\mu\ra= \sigma |\sigma\mu\ra$ so that due to the contribution from all the $l_m$s the ground state subspace of $H_{TC}$ splits with energies $-1$ and $1$, and therefore there is no "topological" protection of the state, and it is not useful as a quantum memory anymore. It is worth noting that we can also make the system have a gapless like spectrum, by favoring a situation with a x-particle, or a domain wall present. This excitation can move around at a low energetic cost. Let us assume we have periodic boundary conditions of length $2L_1$, and take
\begin{eqnarray}
H=H_{TC}-{2\over L_1}\sum_{m=1}^{L_1} (l_m+1)-{2\over L_1}\sum_{m=L_1+1}^{2L_1} (1-l_m)
\end{eqnarray}
Note that $|\sigma\mu\ra$ are eigenstates of the system with eigenvalues $-4$. This system can be diagonalized exactly since all terms commute.  If we introduce two $x$ particles in the system in the form:
$|X,\sigma\mu\ra=\prod_{j\in c}\sigma^x_j|\sigma\mu\ra$, where $c$ is a cut from site $1$ to site $L_1$. We now note that: $(1+l_m)|X,-\mu\ra=2|X,-\mu\ra$ if $m\in[1,L_1]$, while $(1-l_m)|X,-\mu\ra=2|X,-\mu\ra$ if $m\in[L_1+1,2L_1]$, since the string $X$ can be commuted with the loop operator $l_mX=-Xl_m$ for $m\in[1,L_1]$. Thus we have:
$H|X,-\mu\ra=(2-8)|X,-\mu\ra=-6|X,-\mu\ra$ this is the new ground state. The cost of moving the end of the string one step is very low: ${4\over L_1}$ if we move in along the $L_1$ direction (and $0$ if we move the string end in the $L_2$ direction). Therefore, as $L_1\rightarrow\infty$ the spectrum becomes gapless.

%\subsection{Thin, but not very very thin}

\subsection{Torus}

As we have seen, the thin torus limit may be unstable to small perturbation. However, the situation is different if we take $L_1$ and $L_2$ to the thermodynamic limit together. We find that if the ratio of $L_2$ and $L_1$ doesn't blow up exponentially as we take that limit, then stability is retained. The idea we will use in order to show this property is to work in each of the four sectors (given the values of $T_1,T_2$) separately, and then compare the energy shifts and the gaps in each sector.

We now compute the expressions for the energy and gap in each subspace as computed using the cluster expansion formalism from 
\begin{eqnarray}
Z_{\sigma\mu}(N)=\la\sigma \mu |e^{-NH}|\sigma \mu\ra
\end{eqnarray}

To use the cluster expansion in this case we note that any cluster $c$ containing a full loop around the torus, when expanded in the interactions $V_I$ may contain loop operators. For general perturbations involving $\sigma_z$ and $\sigma_x$ operations, there will be no factorizability property of a pair of loops, even if they are arbitrarily far along the torus. This is clear from the following simple example:
\begin{eqnarray}
1=\la\sigma \mu |\prod_{i=1}^{L_1}\sigma^x_{i1}\prod_{i=1}^{L_1}\sigma^x_{ik}|\sigma \mu\ra\neq\la\sigma \mu |\prod_{i=1}^{L_1}\sigma^x_{i1}|\sigma \mu\ra\la\sigma \mu |\prod_{i=1}^{L_1}\sigma^x_{ik}|\sigma \mu\ra=0,
\end{eqnarray}
where $1\leq k\leq L_1$ is an arbitrary location for the second loop. The left hand side of this equation reflects that the loops $\prod_{i=1}^{L_1}\sigma^x_{i1}$ and $\prod_{i=1}^{L_1}\sigma^x_{ik}$ are topologically equivalent. While the right hand side holds since $$\la\sigma \mu |\prod_{i=1}^{L_1}\sigma^x_{ik}|\sigma \mu\ra=\la\sigma \mu |T_2\prod_{i=1}^{L_1}\sigma^x_{ik}T_2|\sigma \mu\ra=\la\sigma \mu |\prod_{i=1}^{L_1}\sigma^x_{ik}(-T_2)T_2|\sigma \mu\ra=-\la\sigma \mu |\prod_{i=1}^{L_1}\sigma^x_{ik}|\sigma \mu\ra.$$

The crucial simplification when using only $\sigma_z$ perturbations, is that any such loop will act on the ground states $|\sigma\mu\ra$ as a scalar and will commute with all other operators appearing in a cluster $c$, therefore factorizability is retained. We may now use the cluster expansion as derived in the previous section. 

However, we have to consider separately the subspaces with given numbers of $x$ particles, since the $V_i$ here cannot create those. Thus we consider first:

\subsubsection{Energy shifts in the subspace with no $x$-particles present}
We proceed to compare the expressions for the energy obtained from the cluster expansion for each of the ground states. We write:
\begin{eqnarray}
E_{\sigma\mu}=\sum_{{X{\rm connected\,clusters\,starting\,at\,t=0}}}\omega_{\sigma\mu}(X)
\end{eqnarray}
The $o(e^{-N\gamma})$ term is:
\begin{eqnarray}
\sum_{{X{\rm connected\,clusters\,starting\,at\,t=0\,with\, l(X)>N}}}(l(X)-N)\omega_{\sigma\mu}(X)
\end{eqnarray}
where $\omega_{\sigma\mu}$ is given by \eqref{omegaX} where expectation values are evaluated in the $|\sigma\mu\ra$ state.

We note that for any cluster of support less than a loop length, all ground states will have exactly the same contribution. Thus the difference between $E_{++},E_{--}$ etc, is due to loops of support larger then $min(L_1,L_2)$. The contribution from such loops is of order exponential in the torus length i.e. the difference between say $E_{+-}$ and $E_{++}$ is $L_1o(e^{-\kappa L_2})$ for some $\kappa$, where the $L_1$ factor comes from the number of possible starting points at $t=0$. Since these states remained gapped from the rest of the spectrum by the same arguments as we had for the sphere, we conclude that the four new lowest energies are ensured to be exponentially close in energy if we demand that $L_ie^{-\kappa L_j}\rightarrow 0$ when $L_i,L_j\rightarrow \infty$. 

\subsubsection{Energy when x-particles are present}

To complete the argument we must now consider the rest of the spectrum: Indeed, the $V_I$ perturbations we introduced can only create one kind of particle, and so the considerations above only apply to the cyclic subspace containing no $x$ particles. Here we have to show that the energetic penalty for introducing $x$ particles, is larger then energy shifts in this subspace compared to the previous section (and contrary to the situation shown above for the thin torus). This can be addressed in the following way: we check the behavior of $e^{-NH}$ on states containing a fixed number of $x$ particles. 

To do so we introduce the $x$ particles by applying to one of the ground states
\begin{eqnarray}
X_C=\prod_{\alpha\in C}\sigma^x_{\alpha},
\end{eqnarray}
where $C$ is some set of edges.
Let us denote by $n_x$ the number of $x$ particles associated with $C$. Since $X_C^2=1$ and $X_C^{\dag}=X_C$ it holds that:
\begin{eqnarray}
\la X_Ce^{-NH}X_C\ra=\la e^{-NX_CHX_C}\ra.
\end{eqnarray}
$X_C$ is a unitary transformation that acts as $X_C\sigma^z_{\alpha}X_C=-\sigma^z_{\alpha}$ if $\alpha\in C$ and $X_C\sigma^z_{\alpha}X_C=\sigma^z_{\alpha}$ otherwise. Therefore $X_CHX_C$ is a Hamiltonian where some of the $\sigma_z$ operators have changed sign, and some of the $P_{\square}$ terms became $P_{\square}^{\perp}=1-P_{\square}$. These terms still commute with the rest of the Hamiltonian, and have eigenvalue $1$ when applied to a state not containing $x$ particles. The effect of $X_C$ can therefore be summarized as:
\begin{eqnarray}\label{XC}
\la X_Ce^{-NH}X_C\ra=\la e^{-N(H_C+n_x)}\ra
\end{eqnarray}
where $H_C=H_{TC}+\sum X_C V_I X_C$.

We may compute the cluster expansion associated with the new Hamiltonian $H_C$. We have to identify the clusters which may 
feel the sign changes due to $\sigma_z$ sign changes in some of the $V_i$s. We note that the only clusters where a given $\sigma_z$ appears an odd number of times (and so will be sensitive to the sign change) and which have non zero expectation in the ground state of $H_{TC}$ are those which contain closed loops (so they do not create particles). Such clusters are either encircling the entire torus, and so have weight $o(e^{-\kappa L})$, or have to go around one of the $x$ particles. The contributions from clusters encircling a particular $x$ charge can be bounded by $\sum_{d=1}^{\infty} c_1 d  e^{-\kappa d} <c_2e^{-\kappa}$ (This is equivalent to stating that the energy shift associated with a particular site is small when the perturbations are weak, or if the perturbation is translation invariant, the energy shift per unit volume is small). Thus if we demand that $c_2e^{-\kappa}<\epsilon$, we have that 
$|E_0(H)-E_0(H_C)|<\epsilon n_x+o(e^{-\kappa L})$, therefore, it follows from \eqref{XC} that 
\begin{eqnarray}
{\log(\la X_Ce^{-NH}X_C\ra)-\log(\la e^{-NH}\ra)\over N}\geq n_x(1-\epsilon)+o(e^{-\kappa L})
\end{eqnarray}
as $N\rightarrow\infty$ and we see that the states with $x$ particles excited are separated by a gap of order $n_x(1-\epsilon)+o(e^{-\kappa L})$ from the lowest states not containing $x$ particles.

\section{Summary}
In this paper we explored the role of perturbations on the behavior of certain topological lattice models. To simplify the treatment we started at the outset from the assumption that the states are described by braid diagrams, with the property that any two non-linked braids factorize. We have proved that for weak enough perturbations, a topological phase defined by an anyon lattice on a sphere is stable.
However, one may expect that more general systems may also be treated similarly. For systems on a torus we confined ourselves to the case of $z$ perturbations of the toric code. Here we have shown how the stability of the model depends on the asymptotic ratio of the two torus directions. As expected, the system is stable whenever the ratio is not exponential. We expect such a result to hold for arbitrary local perturbations which are weak enough.
This is the subject of an upcoming work.

\section*{Acknowledgment}
It is a pleasure to thank M. Hastings for introducing the problem to me, as well as numerous helpful suggestions. I would also like to thank M. Freedman,  A. Hamma, K. Shtengel and F. Verstraete and for discussions.
\bibliographystyle{elsart-num}
\bibliography{/Users/iklich/Documents/Work/KlichBib}
\end{document}